\documentclass[pra,twocolumn,letterpaper,showpacs]{revtex4}

\usepackage{amssymb}
\usepackage{epsfig}

\usepackage{amsfonts}
\newcommand{\field}[1]{\mathbb{#1}}

\newcommand{\R}{\field{R}}


\begin{document} 

\title{Cooling ultracold bosons in optical lattices by spectral transform}

\author{David~L.~Feder} 
\affiliation{Department of Physics and Astronomy and Institute for Quantum
Information Science, University of Calgary, Calgary, Alberta, Canada T2N 1N4}

\date{\today}

\begin{abstract}
It is shown theoretically how to directly obtain the energy distribution of a 
weakly interacting gas of bosons confined in an optical lattice in the 
tight-binding limit. This is accomplished by adding a linear potential to a 
suitably prepared lattice, and allowing the gas to evolve under the influence 
of the total potential. After a prescribed time, a spectral transform is
effected where each (highly non-local) energy state is transformed into a 
distinct site of the lattice, thus allowing the energy distribution to be 
(non-destructively) imaged in real space. Evolving for twice the time returns 
the atoms to their initial state. The results suggest efficient methods to 
both measure the temperature in situ, as well as to cool atoms within the 
lattice: after applying the spectral transform one simply needs to remove 
atoms from all but a few lattice sites. Using exact numerical calculations, 
the effects of interactions and errors in the application of the lattice are 
examined.
\end{abstract}

\maketitle

\section{Introduction}
\label{sec:introduction}

Despite the strong interest in the properties of ultracold bosons confined in
optical lattices~\cite{Bloch05,Morsch06,Lewenstein07,Greiner08,Bloch08}, the 
accurate 
determination and efficient control of the temperature remains a persistent 
problem. The initial loading of the atoms into optical lattices can cause
significant heating if it is not performed adiabatically~\cite{Gericke07}
which becomes increasing difficult as the lattice deepens~\cite{Sklarz02}.
Likewise, adiabatic changes to the lattice potential itself can both increase
or decrease the temperature~\cite{Blakie04,Rey06,Blakie07}. Instrinsic 
problems such as laser motion~\cite{Schmid06}, atomic motion~\cite{Scott04}, or
impurities~\cite{Catani08,Cramer08} can also adversely affect the temperature.
The inability to achieve very low temperatures in optical lattices is one of
the primary obstacles to achieving and controlling interesting quantum states 
in these systems~\cite{Moritz03,Paredes04,Reischl05}.

There is currently no efficient way to directly measure the temperature of the 
bosons when quantum mechanical tunneling within the lattice is appreciable. 
This weakly correlated regime is applicable for sufficiently shallow lattices, 
weak interactions, or large numbers of atoms at each site~\cite{Cirac03}. The 
standard method is to release the atoms from the trap, and is fully 
destructive~\cite{Lin08}. The shadow-image of 
the fully expanded cloud then corresponds approximately to the momentum 
distribution in the trap, though the presence of atomic interactions modifies 
this simple picture somewhat~\cite{Diener07,Gerbier07}. In contrast, in the 
weak tunneling (strongly correlated) limit there are several proposed methods 
for inferring the temperature, including measuring the density distribution of 
doubly occupied sites~\cite{Pupillo06} and the atomic number 
fluctuations~\cite{Capogrosso07}.

Unfortunately, directly cooling weakly correlated atoms in an optical lattice 
is not straightforward. Only one procedure has been proposed, based on a 
combination of sophisticated quantum optics and many body 
effects~\cite{Griessner06}. In contrast, under conditions of weak tunneling 
several techniques have been proposed~\cite{Popp06}. The standard evaporative 
cooling technique for bosons in harmonic traps~\cite{Pethick01} is predicated 
on the correspondence between the energy and spatial extent of the 
single-particle quantum states. Atoms far from the trap center tend to have a 
higher energy, so their gradual removal combined with rethermalization can 
rapidly cool the particles. This approach unfortunately cannot be applied to 
optical lattices even if they are shallow, however, because all 
single-particle energy states span the length of the lattice. 

This work proposes that optical lattices can be suitably prepared in order to
allow for the direct measurement of the temperature of a weakly correlated Bose 
gas, and also to efficiently cool it. The main idea is that if one could 
vary the depth of the optical lattice quadratically along its length, then to
an excellent approximation the low-energy single-particle states of the system 
are the discrete-space analogs of the harmonic oscillator Hermite polynomials, 
known as Krawtchouk functions~\cite{Krawtchouk29,Alvarez05}. As discussed in
detail below, this could be accomplished using at least two distinct and
readily available methods. First, one could change the focal length of one of 
the optical lattice lasers, so that the intensity maxima of the two beams are 
displaced relative to one another. Second, one could add an additional optical
lattice at a small angle relative to the first. In both cases, if the 
temperature is already relatively cold, then an {\it in-situ} non-destructive 
image will reveal a bimodal density distribution much like those found for 
bosons in harmonic traps~\cite{Miesner98}. Evaporative cooling can then be 
performed directly. 

For larger temperatures where the tight-binding band is close to 
full~\cite{Greiner01} and all single-particle states are strongly overlapping, 
the energy distribution of the gas can be obtained by performing a full 
spectral transformation. This is accomplished by adding a small linear 
potential~\cite{Greiner02}, and allowing the system 
to evolve in time. Single-particle states with low energies will evolve with 
high probability into the occupation of lattice sites on the left, while 
high-energy states will migrate to sites on the right. Thus, after a prescribed
evolution time the spatial profile of the atomic density yields the initial
energy distribution. Further cooling can then be effected simply by 
preferentially removing atoms toward the right by imaging with a resonant 
laser. Time-evolving further performs the reverse transformation, mapping
the single-particle states back to their original representation.

While this approach is exact in the limit of non-interacting particles (and
therefore much in the spirit of evaporative cooling ideas in harmonic traps)
and only for a particular choice of tight-binding Hamiltonian, it is robust 
with respect to both atomic interactions and to errors in the application of 
the optical lattice, as shown in detail toward the end of this manuscript. It 
is equally applicable to one, two, or three dimensional systems.

The manuscript is organized as follows. The basic formalism for atoms in
optical lattices and the tight-binding approximation are reviewed in
Sec.~\ref{sec:optlat}. The conditions under which it is possible in principle
to perform a spectral transform are discussed in Section~\ref{sec:transform}.
The Krawtchouk functions are introduced in Section~\ref{sec:kraw}, and the 
condition under which these may effect a spectral transform is also shown.
The robustness of the results to the presence of interactions and 
approximations to the optical lattice potential are explored in 
Section~\ref{sec:approx}, and the results are discussed and summarized in 
Section~\ref{sec:conclusions}.

\section{Atoms in Optical Lattices}
\label{sec:optlat}

Optical lattices are standing waves formed by overlapping, focused lasers. 
In the case of a one-dimensional (1D) lattice, the lasers can be considered to 
be counterpropagating. Assuming that the lattice is oriented along the axial 
$\hat{z}$ direction and is axisymmetric, the lattice potential 
is~\cite{Chiofalo00}
\begin{equation}
V_{\rm l}(r,z)=-V(z)\exp(-r^2/r_0^2)\sin^2(kz),
\end{equation}
where $r$ is the transverse coordinate, $V(z)>0$ is the lattice depth that
may depend on the axial coordinate, $r_0$ is the transverse beam radius, and 
$k=2\pi/\lambda$ is the wavenumber of the light with wavelength $\lambda$. 
This expression assumes that the confinement is strongest (the potential is 
most negative) at points of largest laser intensity (i.e.\ when the frequency 
$c/\lambda$ of the laser is smaller 
than that of the main atomic resonance, known as red-detuning~\cite{Pethick01}).
Allowing for the possibility of an additional external potential $V_e(z)$ 
aligned along the lattice axis, the Hamiltonian for interacting bosons is 
\begin{eqnarray}
H&=&\int d^3r\psi^{\dag}(r,z)\left[-\frac{\hbar^2}{2m}\nabla^2
+V_{\rm l}(r,z)+V_{\rm e}(z)\right]\psi^{\vphantom{\dag}}(r,z)\nonumber \\
&+&\frac{g}{2}\int d^3rd^3r'\psi^{\dag}(r,z)
\psi^{\dag}(r',z')\psi^{\vphantom{\dag}}(r',z')
\psi^{\vphantom{\dag}}(r,z),\hphantom{aa}
\end{eqnarray}
where $\psi^{\dag}(r,z)$ creates a boson field at position $(r,z)$, 
$g=4\pi\hbar^2a/m$ is the two-body coupling constant~\cite{Pethick01},
and $a$ is the $s$-wave scattering length. For $^{87}$Rb, for example, the
value of the scattering length is $a\approx 100a_0=5.29$~nm~\cite{vanKempen02},
with $a_0$ the Bohr radius. It is convenient to express all energies in terms
of the atomic recoil energy $E_R\equiv\hbar^2k^2/2m=2\pi^2\hbar^2/m\lambda^2$. 
The coupling constant is then $g=(2\lambda^2a/\pi)E_R$.

In the limit of very strong transverse confinement~\cite{Strecker02}, all of 
the boson fields will occupy the lowest-energy eigenstate of the confining 
potential in the transverse direction. Neglecting the possible $z$-dependence 
of the lattice potential for the time being, one obtains 
$V_{\rm l}(r)\approx -V(1-r^2/r_0^2)=-V+Vr^2/r_0^2$
which is approximately harmonic with effective frequency 
$\omega=\sqrt{2V/mr_0^2}$ and harmonic oscillator length
$\ell=\sqrt{\hbar/m\omega}$. Making the substitution 
$\psi^{\dag}(r,z)\approx\sqrt{2/\pi\ell^2}\exp(-r^2/2\ell^2)
\psi^{\dag}(z)$ and integrating over the radial coordinate gives a rescaled
coupling constant $g_{\rm 1D}=g/\pi\ell^2=4\pi a\hbar\omega
=4\sqrt{\overline{V}}(a\lambda/r_0)E_R$, where $\overline{V}=V/E_R$ is the 
lattice depth in recoil energies. 

Focusing the optical lattice lasers generally leads to an axially-dependent
lattice depth $V(z)$ that varies over the Rayleigh length. As a result, the 
lattice depth is usually largest at the center and exponentially decreasing 
on either side $V(z)=V_0\exp(-z^2/z_0^2)$, though with a very long beam waist
$z_0$. Beam waists for optical potentials as long as 
$125~\mu$m~\cite{Meyrath05} and for optical lattices as small as 
$60~\mu$m~\cite{Nelson07} have been employed in recent experiments. 

An excellent description of ultracold atoms in optical lattice potentials of
the moderate depth discussed above is provided by the Bose-Hubbard
Hamiltonian~\cite{Jaksch98}. The Bose field operators are expanded 
$\psi(z)=\sum_i b_iw(z-z_i)$ in a basis of Wannier functions $w(z-z_i)$ that 
are highly localized near a lattice site $i$. Assuming that the energies 
related to the particle dynamics and the temperature $k_BT$ are both small 
compared with the bandgap, one obtains the Bose-Hubbard 
Hamiltonian~\cite{Fisher89}:
\begin{equation}
H=-\sum_{\langle ij\rangle}J_{i,j}b_i^{\dag}b_j^{\vphantom{\dag}}
+\frac{U}{2}\sum_in_i(n_i-1)+\sum_i\varepsilon_ib_i^{\dag}
b_i^{\vphantom{\dag}},
\label{Hubbard}
\end{equation}
where $\langle ij\rangle$ denotes a sum over nearest neighbours (i.e.\ 
$j=i\pm 1$), $b_i^{\dag}$ creates a boson at site $i$, and
$n_i=b_i^{\dag}b_i^{\vphantom{\dag}}$ is the density operator for site $i$.
For sufficiently deep lattices $V_0\gtrsim 5E_R$, one need only consider
tunneling of atoms between nearest-neighbor sites and interactions of atoms 
within the same lattice site; the amplitudes for hopping between 
next-nearest-neighbor sites and the strength of interactions between atoms on 
neighboring sites are smaller by at least an order of magnitude, 
respectively~\cite{Jaksch98}. 

If the external potential $V_e(z)$ is slowly varying over the lattice spacing, 
the hopping coefficient is 
\begin{equation}
J_{i,j}=\int dz w^*(z-z_i)\left[-\frac{\hbar^2}{2m}\nabla^2
+V_{\rm l}(z)\right]w(z-z_j).
\label{hop}
\end{equation}
In the very deep lattice limit where the value of the lattice depth at site $i$
(in recoil energies) is large $\overline{V}_i\equiv V_i/E_R\gg 1$, the Wannier 
functions can be approximated by the ground-state Gaussian wavefunctions for 
each separate well $w(z)\approx e^{-z^2/2d^2}/\sqrt{d\sqrt{\pi}}$ with
$d=(k\overline{V}_i^{1/4})^{-1}$. Then the hopping amplitudes~(\ref{hop}) are 
given by $J_{i,j}/E_R=(\pi^2/4)\overline{V}_i\exp(-\pi^2\sqrt{\overline{V}_i}
/4)$.
A better approximation for the Wannier functions instead 
gives~\cite{Bloch08,Pupillo06}
\begin{equation}
{J_{i,j}\over E_R}\approx\sqrt{16\over\pi}\overline{V}_i^{3/4}
\exp\left(-2\sqrt{\overline{V}_i}\right).
\label{tunneling}
\end{equation}
Meanwhile, the on-site interaction term is
\begin{equation}
U=g_{\rm 1D}\int dz |w(z)|^4.
\end{equation}
Again approximating the Wannier functions by harmonic oscillator functions, one
obtains
\begin{equation}
{U\over E_R}=\sqrt{8\over\pi}{a\lambda k\over r_0}\overline{V}^{3/4}
=\sqrt{32\pi}{a\over r_0}\overline{V}^{3/4}.
\label{Uapprox}
\end{equation}
Finally, the site-dependent local energy is 
\begin{equation}
\varepsilon_i=\int dz w^*(z-z_i)\left[-\frac{\hbar^2}{2m}\nabla^2
+V_{\rm l}(z)+V_{\rm e}(z)\right]w(z-z_i).
\label{e0}
\end{equation}
This latter term is a constant energy offset for homogeneous lattice potentials
and no exernal potential, and is usually dropped. In the present case, 
however, the lattice amplitude will vary slowly over the lattice spacing 
$\lambda/2$. This will give rise to a spatial variation of $\varepsilon_i$, 
which for large lattices is equivalent to a smoothly varying external 
potential even in the absence of any $V_{\rm e}(z)$. Again using the
local harmonic oscillator solutions for the Wannier functions, one obtains
$\varepsilon_i/E_R\approx\sqrt{\overline{V}_i}$.

The onset of strong correlations for bosons in quasi-1D optical lattices with 
an average of $\langle n\rangle\gg 1$ atoms per site occurs when
$U/J>2.2\langle n\rangle$~\cite{Hamer79,Roomany80,Bloch08}. Combining 
Eqs.~(\ref{tunneling}) and (\ref{Uapprox}), one obtains the criterion
\begin{equation}
\sqrt{2}\pi{a\over r_0}e^{2\sqrt{\overline{V}}}>2.2\langle n\rangle.
\end{equation}
The value of $r_0$ is set by imposing the quasi-1D criterion 
$\hbar\omega\gtrsim\overline{V}E_R$, or 
$r_0/\lambda\ll(\pi\sqrt{\overline{V}})^{-1}$. This condition is in fact much
more restrictive than necessary because at low temperatures $T\propto J$ 
(for example, the BEC transition temperature for a 3D lattice is 
$k_BT_c=5.59J$~\cite{Bloch08}) only the first band is occupied, and the 
bandwidth in the 1D tight binding limit is $w=2J\ll V$. With 
$\hbar\omega=\overline{V}E_R$, $a=5.29$~nm, and $\lambda=800$~nm, and a 
typical value $\langle n\rangle=1000$ one obtains 
$\overline{V}_{\rm max}\lesssim 16$ and $r_0\approx 32$~nm to ensure that the
atoms remain in the weakly correlated regime. This transverse length scale 
corresponds to a confining frequency $\omega/2\pi\approx 100$~kHz, a typical 
value in current experiments~\cite{Bloch08}. With $\langle n\rangle=10\,000$
instead, one obtains $\overline{V}_{\rm max}\approx 25$. 

Far in the weakly correlated regime, the amplitude for nearest-neighbor 
tunneling from site to site tends to be larger (and therefore more important) 
than the strength of particle interactions. As a first approximation, we may 
therefore set $U=0$; the consequences of $U\neq 0$ will be explored in 
Sec.~\ref{sec:approx}. Furthermore, it will be assumed that $V_e(z)$ can be 
chosen in such a way to ensure that $\varepsilon_i$ is approximately constant,
so that it can be ignored (this assumption will be further justified in 
Sec.~\ref{quadratic}). The resulting non-interacting tight-binding Hamiltonian 
can then be written 
\begin{equation}
H_0=-\sum_{\langle ij\rangle}J_{i,j}|i\rangle\langle j|,
\label{H0}
\end{equation}
where the site state kets are defined by 
$|i\rangle\equiv b_i^{\dag}|\mathbf{0}\rangle$ and $|\mathbf{0}\rangle$ is the 
vacuum state.

In principle, the position indices $i$ and $j$ are unrestricted, but in
practice the lattice has a finite number of occupied sites. This could arise
naturally because of the inherent spatial variation of the lattice potential
discussed above, in which larger number of atoms will occupy the central
region where the potential is deepest. To enforce occupation of a specific
part of the lattice, one can apply `end-cap' potentials formed from tightly
focused blue-detuned lasers~\cite{Strecker02,Meyrath05}. In this case, one
can effectively enforce the `hard-wall' boundary conditions $i\in\{0,N\}$ with 
$j=i\pm 1$ when $0<i<N$, but $j=1$ only for $i=0$ and $j=N-1$ only when $i=N$.

\section{Spectral Transform}
\label{sec:transform}

\subsection{Introduction}

As is well-known from elementary quantum mechanics, an arbitrary quantum state 
$|\psi(j)\rangle$ defined on $N+1$ discrete sites labelled by index $j$ can 
always be expressed in terms of an arbitrary basis of dimension $N+1$
\begin{equation}
|\psi(j)\rangle=\sum_{n=0}^N c_n|\varphi_n(j)\rangle,
\label{wave}
\end{equation}
where the $c_n$ are generally complex coefficients. The set of states 
$\{|\varphi_n(j)\rangle,n\in 0,N\}$ is assumed to be both orthonormal
\begin{equation}
\sum_{j=0}^N\langle\varphi_m(j)|\varphi_n(j)\rangle\equiv\langle\varphi_m|
\varphi_n\rangle=\delta_{m,n}
\label{ortho}
\end{equation}
and complete
\begin{equation}
\sum_n\langle\varphi_n(j)|\varphi_n(k)\rangle=\delta_{j,k},
\label{complete}
\end{equation}
where the Kronecker delta is defined as $\delta_{j,k}=1$ only if $j=k$ and is
zero otherwise. Multiplying both sides of Eq.~(\ref{wave}) on the left by
$\langle\varphi_m(j)|$ and summing over $j$, one obtains the coefficients using 
(\ref{ortho}):
\begin{equation}
c_n=\sum_j\langle\varphi_n(j)|\psi(j)\rangle\equiv\langle\varphi_n|\psi\rangle.
\end{equation}
One could have obtained this directly simply by inserting the complete set, 
$|\psi\rangle=\sum_n|\varphi_n\rangle\langle\varphi_n|\psi\rangle$.

The relevance of the above discussion to the present work is that the 
coefficients $c_n$ can be considered as the representation of the original
wavefunction in the `spectral basis' rather than the original site basis. While
the elements of the original wavefunction in the site basis are
\begin{equation}
\langle j|\psi\rangle\equiv\psi(j)=\sum_nc_n\langle j|\varphi_n\rangle
=\sum_nc_n\varphi_n(j),
\label{trans1}
\end{equation}
the elements of the same wavefunction in the spectral basis are
\begin{equation}
\langle \varphi_n|\psi\rangle\equiv c_n=\sum_j\langle\varphi_n|j\rangle\langle 
j|\psi\rangle=\sum_j\psi(j)\varphi^*_n(j).
\label{trans2}
\end{equation}
Eqs~(\ref{trans1}) and (\ref{trans2}) are transforms of each other through the
basis functions $\varphi_n(j)$; the second can be immediately obtained from the
first from their orthonormality~(\ref{ortho}), and the converse follows from 
their completeness~(\ref{complete}). These generalize the Fourier transforms 
that are specific to the regular lattice case, in which case 
$\varphi_n(j)\propto\exp[i2\pi nj/(N+1)]$.

The purpose of the current work is to show that the coefficients $c_n$ in 
Eq.~(\ref{wave}) can be obtained directly in experiments without having to
explicitly carry out the transformation~(\ref{trans2}). In particular, if one
could implement the spectral transform 
\begin{equation}
|\varphi_n\rangle\to e^{i\phi_n}|n\rangle
\label{transdes}
\end{equation}
with arbitrary phases $\phi_n$, then all particle densities would be 
transformed as $|\psi(j)|^2\to|c_j|^2$:
\begin{eqnarray}
|\psi(j)|^2&=&\langle j|\psi\rangle\langle\psi|j\rangle
=\sum_nc_n\langle j|\varphi_n\rangle\sum_mc_m^*\langle\varphi_m|j\rangle
\nonumber \\
&\to&\sum_{m,n}c_nc_m^*e^{i(\phi_n-\phi_m)}\langle j|n\rangle\langle m|j\rangle
=|c_j|^2.
\end{eqnarray}
Thus after the spectral transformation, the magnitude of the coefficients in 
the expansion~(\ref{wave}) would be obtained simply from the values of the 
density at each lattice point.

\subsection{Transform by Hamiltonian Evolution}

Consider a single particle hopping on $N+1$ sites of a 1D lattice with 
hard-wall boundary conditions, as defined by the non-interacting tight-binding
Hamiltonian~(\ref{H0}) with $J_{i,j}=J_{j,i}$ and $J_{-1,0}=J_{N,N+1}=0$.
This Hamiltonian has $N+1$ eigenvectors $|v_j\rangle$ with eigenvalues 
$\epsilon_j$. Now consider the perturbed Hamiltonian $H=H_0+H_1$ with
completely local (i.e.\ on-site) potential
\begin{equation}
H_1=\sum_{j=0}^NV_j|j\rangle\langle j|
\label{potential}
\end{equation}
with real $V_i$. (Note that the $V_i$ is an external local potential that is 
applied in {\it addition} to the $V_e(z)$ chosen above to ensure constant
$\varepsilon_i$). The objective is to determine the $V_j$ for a given choice
of $J_{i,j}$ such that time-evolution governed by the full Hamiltonian $H$ 
effects the desired spectral transformation~(\ref{transdes}), but now with
$|v_n\rangle$ taking the place of the $|\phi_n\rangle$ above:
\begin{equation}
\exp(-iHt_s/\hbar)|v_j\rangle=e^{i\phi_j}|j\rangle,
\label{transform}
\end{equation}
where $t_s$ is some fixed time satisfying the condition for any $j\in\{0,N\}$
and $\phi_j$ is a possible arbitrary phase.

It is insightful to write the desired unitary $U=\exp(-iHt_s/\hbar)$ in its 
spectral decomposition
\begin{equation}
U=\sum_{k=0}^N\exp(-i\epsilon_k't_s/\hbar)|v_k'\rangle\langle v_k'|,
\end{equation}
where $\epsilon_k'$ and $|v_k'\rangle$ are the eigenvalues and eigenvectors of
$H$, respectively. The transform~(\ref{transform}) can then be written
\begin{equation}
U|v_j\rangle=\sum_{k=0}^N\exp(-i\epsilon_k't_s/\hbar)|v_k'\rangle\langle v_k'
|v_j\rangle=e^{i\phi_j}|j\rangle.
\label{transform2}
\end{equation}
Clearly, if 
\begin{equation}
\langle v_k'|v_j\rangle=\exp(i\epsilon_k't_s/\hbar)e^{i\phi_j}\langle v_k'
|j\rangle
\end{equation}
then one would guarantee that 
\begin{equation}
U|v_j\rangle=e^{i\phi_j}\left(\sum_{k=0}^N|v_k'\rangle\langle v_k'|\right)
|j\rangle=e^{i\phi_j}|j\rangle
\end{equation}
where the term in parentheses is unity because the eigenbasis of $H$ is 
complete.

It would be preferable if the system would evolve back to the original 
eigenbasis after an additional time $t_s$, at least up to an overall phase. 
This is equivalent to requiring that
\begin{equation}
U^2=\sum_{k=0}^N\exp(-i2\epsilon_k't_s/\hbar)|v_k'\rangle\langle v_k'|=I
e^{i\theta},
\end{equation}
where $I$ is the identity matrix of dimension $N+1$ and $\theta$ is some phase.
This condition is most generally obtained when $2\epsilon_k't_s/\hbar=2\pi n
-\theta$ for any $k,n$. Suppose that we choose $n=k^m$ with $m$ some arbitrary
integer ($m=1$ implies the spectrum is linear, $m=2$ gives a quadratic 
spectrum, etc.). Then $\epsilon_k'=\epsilon_0'+ck^m$, with 
$\epsilon_0'=-\hbar\theta/2t_s$ and $c=\hbar\pi/t_s$. Eq.~(\ref{transform2})
can now be written
\begin{eqnarray}
U|v_j\rangle&=&e^{-i\epsilon_0't_s/\hbar}\sum_{k=0}^Ne^{-ick^mt_s/\hbar}
|v_k'\rangle\langle v_k'|v_j\rangle\label{Uvj} \\
&=&e^{i\theta/2}\sum_{k=0}^Ne^{-i\pi k^m}|v_k'\rangle
\langle v_k'|v_j\rangle\nonumber \\
&=&e^{i\theta/2}\sum_{k=0}^N(-1)^k|v_k'\rangle\langle v_k'|v_j\rangle
=e^{i\phi_j}|j\rangle.\nonumber
\label{transform3}
\end{eqnarray}
Thus, in order to perform a spectral transform that is invertible by 
repetition, one requires
\begin{equation}
\langle v_k'|v_j\rangle=(-1)^ke^{i(\phi_j-\theta/2)}\langle v_k'|j\rangle
\quad\forall k,j\in\{0,N\}.
\label{condition}
\end{equation}
In short, the overlap of the eigenvectors of the full Hamiltonian $H$ with
those of the unperturbed Hamiltonian $H_0$ must be trivially related to their
overlap with the site basis vectors. Furthermore, the perturbation to $H_0$ 
must be local (i.e.\ be strictly an externally applied spatially dependent 
potential), and the spectrum of the full Hamiltonian must be an integer power
in the quantum number. 

Consider the homogeneous optical lattice with $J_{i,j}=J$ 
on $N$ sites. Hard-wall boundary conditions force all functions to zero 
at the sites labeled by $i=-1$ and $i=N+1$. The unperturbed eigenvalues in 
ascending order are $\epsilon_k=-2J\cos[\pi(k+1)/(N+2)]$ with $k\in\{0,N\}$,
and the eigenvectors are similar to Chebyshev polynomials,
$|v_k\rangle=\sqrt{2/(N+2)}\sin\left[{\pi(k+1)(j+1)/(N+2)}\right]$, where $k$ 
is the eigenvalue index and $j\in\{0,N\}$ corresponds to the lattice site. It 
is straightforward to verify that choosing $V_k=\epsilon_k$, i.e.\ where the 
perturbation corresponds to the eigenvalues of the original unperturbed 
Hamiltonian in ascending order, indeed yields perturbed eigenvectors 
$|v_k'\rangle$ that satisfy the desired overlap relation
$\langle v_k'|v_j\rangle=(-1)^{k+j}\langle v_k'|j\rangle$. Unfortunately, the
spectrum of the perturbed Hamiltonian is not of the desired special form
$\epsilon_k'=\epsilon_0'+ck^m$. Conversely, if a perturbation is chosen to 
enforce the eigenvalue criterion, then the overlap criterion cannot be
satisfied. The unfortunate conclusion is that no perturbation to a uniform
optical lattice can be chosen in order to effect a spectral transform. The 
remainder of this work is devoted to showing that both of these conditions 
can be nevertheless be simultaneously satisfied with atoms in specially 
prepared optical lattices.

\section{Krawtchouk Lattice}
\label{sec:kraw}

Consider the site-dependent tunneling amplitudes 
\begin{equation}
J_{i,i+1}\equiv J\sqrt{(i+1)(N-i)},
\label{wire}
\end{equation}
where $i\in\{0,N-1\}$ and $J$ is a bare tunneling coefficient with units 
of energy. The associated weighted lattice Hamiltonian has been recently 
investigated in the context of quantum 
walks~\cite{Shenvi03,Christandl04,Christandl05,Feder06}, the quantum 
mechanical generalization of the classical 
random walk~\cite{Kempe03,Ambainis03}. In particular, with these coefficients 
in the governing Hamiltonian, a quantum state initialized in the left-most 
lattice site will evolve with perfect probability to the rightmost lattice 
site after a fixed amount of time. For this reason, this Hamiltonian and its
generalizations have been denoted a perfect quantum wires~\cite{Fitzsimons06}. 
Propagating further for the same amount of time returns the state to the first
site. Likewise, states at any site will be transferred to sites symmetrically 
located around the lattice center, so time-evolution effects a perfect 
reflection of any state about the midpoint. Much of the interest in this and
related systems stems from their potential uses for quantum computation and
communication, because of their ability to perfectly distribute quantum states 
and entanglement~\cite{Kostak07,DAmico07}.

In the present context the Hamiltonian~(\ref{H0}) with tunneling 
coefficients~(\ref{wire}) is useful 
because it possesses both a linear spectrum and exhibits periodic behavior. 
Indeed, the eigenfunctions turn out to be Krawtchouk functions, which are the 
discrete-space analogs of the Hermite polynomials. Furthermore, it will
be shown below that perturbing the quantum wire by a linear potential will
yield a full Hamiltonian whose states satisfy the criterion~(\ref{condition}), 
thereby yielding a full spectral transformation between energy and coordinate 
space. 

\subsection{Unperturbed Hamiltonian}
\label{sec:H0}

The spectrum of the unperturbed Hamiltonian~(\ref{H0}) is obtained by solving
$H_0|v_k\rangle=\epsilon_k|v_k\rangle$ 
Expanding the eigenfunctions in the site basis 
$|v_k\rangle\equiv\sum_ja_j^{(k)}|j\rangle$ with $a_j^{(k)}$ some unknown 
expansion coefficients, one can write the eigenvalue equation as
\begin{eqnarray}
&&\sum_{i,j}a_j^{(k)}\Big(J_{i+1,i}|i+1\rangle\langle i|j\rangle
+J_{i-1,i}|i-1\rangle\langle i|j\rangle\nonumber \\
&&\qquad\qquad-\epsilon_k|j\rangle\Big)=0,
\end{eqnarray}
which can be conveniently rewritten as the three-term recurrence relation
\begin{equation}
J_{j,j-1}a_{j-1}^{(k)}+J_{j,j+1}a_{j+1}^{(k)}-\epsilon_ka_j^{(k)}=0.
\end{equation}
In terms of the quantum wire tunneling amplitudes~(\ref{wire}) this becomes
\begin{eqnarray}
&&\sqrt{j(N-j+1)}a_{j-1}^{(k)}+\sqrt{(j+1)(N-j)}a_{j+1}^{(k)}\nonumber \\
&&-\tilde{\epsilon}_ka_j^{(k)}=0,
\label{recurrence}
\end{eqnarray}
where the eigenvalues $\tilde{\epsilon}_j\equiv\epsilon_j/J$ are now 
dimensionless.

The recurrence relation~(\ref{recurrence}) defines the normalized Krawtchouk
polynomials $\psi_j^{(p)}(z)$~\cite{Krawtchouk29,Alvarez05}
\begin{eqnarray}
&&\sqrt{j(N-j+1)}\psi_{j-1}^{(p)}(k)+\sqrt{(j+1)(N-j)}\psi_{j+1}^{(p)}(k)
\nonumber \\
&&+\frac{Np+(1-2p)j-k}{\sqrt{p(1-p)}}\psi_j^{(p)}(k)=0,
\label{kraw_recurrence}
\end{eqnarray}
where the Krawtchouk polynomials can be expressed in terms of either
hypergeometric or Jacobi functions:
\begin{eqnarray}
\psi_j^{(p)}(k)&=&\left(\frac{p}{1-p}\right)^{k+j\over 2}(-1)^j(1-p)^{N/2}
\sqrt{\frac{j!(N-j)!}{k!(N-k)!}}\nonumber \\
& &\qquad\times\left(\matrix{N\cr j\cr}\right)\vphantom{F}_2F_1
\left(-j,-k;-N;\frac{1}{p}\right),\nonumber \\
&=&\left(\frac{p}{1-p}\right)^{k+j\over 2}(-1)^j(1-p)^{N/2}
\sqrt{\frac{j!(N-j)!}{k!(N-k)!}}\nonumber \\
& &\qquad\times P_k^{(N-j-k,-N-1)}\left(\frac{2}{p}-1\right),
\label{kraw_def}
\end{eqnarray}
with $p\in\R$ a free parameter. Comparison of Eqs.~(\ref{recurrence}) and 
(\ref{kraw_recurrence}) immediately requires $p=1/2$ to eliminate the $j$ 
(position) dependence of the last coefficient. The spectrum of the unperturbed
Hamiltonian is therefore linear, $\tilde{\epsilon}_k=-N+2k$. The eigenvectors
are 
\begin{equation}
|v_k(j)\rangle={(-1)^j\over 2^{N/2}}\sqrt{\frac{j!(N-j)!}{k!(N-k)!}}
P_k^{(N-j-k,-N-1)}\left(3\right),
\end{equation}
where the position-dependence is now included explicitly.

In the limit of 
large $N$ and small $k$, the Krawtchouk polynomials reduce to Hermite 
polynomials~\cite{Krawtchouk29,Hakioglu00}, reflecting the fact that the 
Hamiltonian for the quantum wire is the discretized version of the harmonic 
oscillator. This is easy to verify directly from the recurrence 
relation~(\ref{recurrence}) in the limit $N\to\infty$ and $j\ll N$:
\begin{equation}
\sqrt{jN}a_{j-1}^{(k)}+\sqrt{(j+1)N}a_{j+1}^{(k)}+(N-2k)a_j^{(k)}=0,
\end{equation}
which is equivalent to the recurrence relation for the normalized Hermite
polynomials $\tilde{H}_j(x)$~\cite{NR}
\begin{equation}
\sqrt{j}\tilde{H}_{j-1}+\sqrt{j+1}\tilde{H}_{j+1}-x\sqrt{2}\tilde{H}_j=0.
\end{equation}
Evidently, $x=\sqrt{2}k$ neglecting the overall constant term. The effective
trapping (oscillation) frequency is therefore $\omega=2J/\hbar$.

\subsection{Full Hamiltonian}

The $j$-dependence of the third coefficient appearing in the recurrence 
relation~(\ref{kraw_recurrence}) indicates that adding a linear potential 
$V_j\propto j$ in (\ref{potential}) to the unperturbed Hamiltonian would still 
yield eigenstates that are Krawtchouk functions. In this case the potential 
would need to be of the form
\begin{equation}
\tilde{V}_j=\frac{1-2p}{\sqrt{p(1-p)}}j,
\label{ramp}
\end{equation}
yielding the softened linear spectrum
\begin{equation}
\epsilon_k'=\frac{1}{\sqrt{p(1-p)}}\left(-Np+k\right)
\label{spectrum}
\end{equation}
and the full expression~(\ref{kraw_def}) for the eigenvectors 
$|v_k'(j)\rangle=\psi_j^{(p)}(k)$.

It remains to be shown that there is a choice of $p$ for the full Hamiltonian 
that can satisfy the main condition~(\ref{condition}). In fact, as will be 
proven below, there is indeed a choice of $p$ that ensures
\begin{equation}
\langle v_k'|v_m\rangle=(-1)^{k+m}\langle v_k'|m\rangle,
\label{condition2}
\end{equation}
which effects the spectral transformation $U|v_m\rangle=(-1)^m|m\rangle$ for 
all $m$, and another choice that performs the mirror-symmetric version 
$U|v_m\rangle=(-1)^m|N-m\rangle$. In terms of the Krawtchouk functions, 
condition~(\ref{condition2}) can be written as
\begin{eqnarray}
&&\sum_{j=0}^N\left(\matrix{N\cr j\cr}\right)^{-1}\frac{(-1)^{k+m}}{2^{N/2}}
\left(\matrix{N\cr m\cr}\right)P_j^{(N-j-m,-N-1)}(3)\nonumber \\
&&\times\left(\frac{p}{1-p}\right)^{j/2}\left(\matrix{N\cr k\cr}\right)
P_j^{(N-j-k,-N-1)}\left(\frac{2}{p}-1\right)\nonumber \\
&&=\left(\frac{p}{1-p}\right)^{m/2}\left(\matrix{N\cr k\cr}\right)
P_m^{(N-m-k,-N-1)}\left(\frac{2}{p}-1\right).\nonumber \\
\label{goal_kraw}
\end{eqnarray}
Eq.~(\ref{goal_kraw}) can be thought of as an $(N+1)$-dimensional matrix
equation $K_1K_2=K_3$, with the sum over $j$ on the left-hand side effecting 
the product of the two Krawtchouk matrices~\cite{Feinsilver01,Feinsilver05} 
\begin{equation}
K_1^{(N)}=\frac{(-1)^k}{2^{N/2}}\left(\matrix{N\cr m\cr}\right)
P_j^{(N-m-j,-N-1)}(3),
\end{equation}
\begin{eqnarray}
K_2^{(N)}&=&(-1)^m\left(\frac{p}{1-p}\right)^{j/2}\left(\matrix{N\cr k\cr}
\right)\nonumber \\
&\times &P_j^{(N-j-k,-N-1)}\left(\frac{2}{p}-1\right),
\end{eqnarray}
which are clearly symmetric under the interchange of row and column indices
$j\leftrightarrow m$ or $j\leftrightarrow k$ respectively, and the right hand 
side given by the symmetric Krawtchouk matrix
\begin{equation}
K_3^{(N)}=\left(\frac{p}{1-p}\right)^{m/2}\left(\matrix{N\cr k\cr}\right)
P_m^{(N-m-k,-N-1)}\left(\frac{2}{p}-1\right).
\end{equation}
The factor $\left(\matrix{N\cr j\cr}\right)$ in the sum~(\ref{goal_kraw}) 
serves as a `metric tensor,' as described in greater detail below. The
Krawtchouk matrix elements are found by fixing $N$ and treating the two
variables in the superscript of the Jacobi polynomials as the row and column
indices, respectively. Thus, the columns of the symmetric Krawtchouk matrices
$K_1^{(N)}$ and $K_2^{(N)}$ correspond to the eigenvalue-ordered eigenvectors
of the unperturbed and perturbed Hamiltonians $H_0$ and $H_1$ on $N+1$ sites,
respectively.

As has been recently discussed in Ref.~\onlinecite{Feinsilver05}, Krawtchouk 
matrices correspond to the `Hamming contraction' of tensor products of 
identical matrices. The hierarchy of Krawtchouk matrices is mathematically 
expressed as $K^{(N)}=r\left\{\left(K^{(1)}\right)^{\otimes N}\right\}$, where 
$K^{(N)}$ is the $N$th order Krawtchouk matrix of dimension $N+1$, and $r$ is 
an operator effecting the Hamming contraction. The contraction is effected by 
labeling the rows/columns of the contracted matrix by the Hamming weight of 
the row/column index (written in binary) for the original matrix, and summing 
all matrix elements whose row (column) indices share the same Hamming weight. 
For example, the first-order Krawtchouk matrix $K_1^{(1)}$ is the Hadamard
operator $\overline{H}$
\begin{equation}
K_1^{(1)}=\frac{1}{\sqrt{2}}\left(\matrix{1 & 1\cr 1 & -1\cr}\right)
=\overline{H},
\end{equation}
and the second-order Krawtchouk matrix $K_1^{(2)}$ is the contraction of the 
tensor product of two Hadamard operators:
\begin{eqnarray}
K_1^{(2)}&=&r\left(\overline{H}\otimes\overline{H}\right)
=r\left[\frac{1}{2}\left(
\matrix{1 & 1 & 1 & 1\cr 
1 & -1 & 1 & -1\cr 
1 & 1 & -1 & -1\cr 
1 & -1 & -1 & 1\cr}\right)\right]\nonumber \\
&=&\frac{1}{2}\left(\matrix{1 & 2 & 1\cr 
2 & 0 & -2\cr 
1 & -2 & 1\cr}\right).
\end{eqnarray}

Equation~(\ref{goal_kraw}) can therefore be interpreted as the product of 
{\it uncontracted} tensor products of first-order Krawtchouk matrices,
\begin{equation}
\left(K_1^{(1)}\right)^{\otimes N}\left(K_2^{(1)}\right)^{\otimes N}
=\left(K_3^{(1)}\right)^{\otimes N},
\end{equation}
which can be alternatively expressed as 
\begin{equation}
\left(K_1^{(1)}K_2^{(2)}\right)^{\otimes N}=\left(K_3^{(1)}\right)^{\otimes N}.
\end{equation}
The `weight factor' $\left(\matrix{N\cr j\cr}\right)$ appearing in the sum 
of Eq.~(\ref{goal_kraw}) can now clearly be understood as accounting for the 
multiplicities for each Hamming weight on the integers between 0 and $2^N$, 
and therefore serves as a `metric' for matrix-matrix multiplications in 
contracted space. Thus, to prove the equality (\ref{goal_kraw}) for arbitrary 
$N$ one need only find the value of $p$ satisfying the $N=1$ case:
\begin{eqnarray}
&&\frac{1}{\sqrt{2}}\left(\matrix{1 & 1\cr 1 & -1\cr}\right)
\left(\matrix{1 & 1\cr \sqrt{\frac{p}{1-p}} & -\sqrt{\frac{1-p}{p}}\cr}\right)
\nonumber \\
&&\qquad=\frac{1}{\sqrt{2}}\left(\matrix{1 & -1\cr
\sqrt{\frac{p}{1-p}} & \sqrt{\frac{1-p}{p}}\cr}\right),
\end{eqnarray}
which yields the unique solution
\begin{equation}
p=\frac{1}{4}\left(2-\sqrt{2}\right).
\end{equation}
It is easy to verify that this solution indeed satisfies Eq.~(\ref{goal_kraw})
for arbitrary choices of $N$. Inserting this result into Eqs.~(\ref{ramp}) and
(\ref{spectrum}) gives the linear potential 
\begin{equation}
\tilde{V}_j=2j
\label{goodpot}
\end{equation}
and the resulting linear spectrum for the perturbed Hamiltonian
\begin{equation}
\epsilon_k'=\sqrt{2}\left(-N+2k\right).
\end{equation}
Because $c=2\sqrt{2}$ in Eq.~(\ref{Uvj}), the time to effect the spectral
transform is 
\begin{equation}
t_s=\frac{\pi}{2\sqrt{2}}\left(\frac{\hbar}{J}\right).
\label{goodtime}
\end{equation}
Eqs.~(\ref{wire}), (\ref{goodpot}) and (\ref{goodtime}) constitute the main 
results of this work.

To perform the mirror-symmetric spectral transform $U|v_m\rangle=|N-m\rangle$
one needs to obtain the value of $p$ that ensures
\begin{equation}
\langle v_k'|v_m\rangle=(-1)^{k+m}\langle v_k'|N-m\rangle,
\label{goal3}
\end{equation}
In this case, one obtains
\begin{eqnarray}
K_3^{(N)}&=&(-1)^{k+m}\left(\frac{p}{1-p}\right)^{m/2}
\left(\matrix{N\cr k\cr}\right)\nonumber \\
&\times&P_{N-m}^{(m-k,-N-1)}\left(\frac{2}{p}-1\right).
\end{eqnarray}
Following the same procedure as discussed above, it is straightforward to show
that to satisfy condition (\ref{goal3}) one requires 
$p=\frac{1}{4}\left(2+\sqrt{2}\right)$. This yields the same prefactor as the
solution above, namely $1/\sqrt{p(1-p)}=2\sqrt{2}$. The spectrum is therefore
the same, but the applied potential now has the opposite sign, 
$\tilde{V}_j=-2j$. Thus, to perform an inverted spectral transform where the
ground state is transformed to the rightmost site rather than the leftmost,
one simply requires the same linear potential but with a negative slope.

\subsection{Discussion of Main Results}

One might be surprised that neither the applied linear 
potential~(\ref{goodpot}) nor the required time~(\ref{goodtime}) depend 
explicitly on the number of lattice sites. Clearly, as $N$ grows, the initial 
wavepacket must traverse an increasing number of lattice sites during the
implementation of the spectral transform. It is important to recall, however,
that the actual tunneling amplitudes~(\ref{wire}) do indeed increase with $N$.
That said, it is not obvious that their values are just what are required to
ensure that $t_s$ remains constant. Consider the time it takes on average to
tunnel between adjacent lattice sites, 
$\overline{t_{i,i+1}}\approx h/J_{i,i+1}$. The total time required for a 
wavepacket to propagate from one end of the lattice to the other is then
approximately 
\begin{equation}
\overline{T}\approx{\hbar\over J}\sum_{i=0}^N{1\over\sqrt{(i+1)(N-i)}}.
\label{sitespan}
\end{equation}
In the large-$N$ limit, one obtains the asymptotic result 
$\lim_{N\to\infty}\overline{T}=(\hbar/J)\pi=h/2J$. This is exactly the period 
for one oscillation of the original Krawtchouk oscillator: the eigenvalues 
$\tilde{\epsilon}_k=2k$ are equivalent to the frequency $2J/\hbar=2\pi/T$, 
which yields $T=h/2J$.

The fact that the Krawtchouk matrices represent contractions of 
tensor products of identical matrices has in fact been anticipated in the 
literature, though in a completely different context. The quantum
wire tunneling amplitudes (\ref{wire}) are obtained by `collapsing' the
hypercube graph to a weighted linear graph, by rendering indistinguishable
vertex labels with the same Hamming weight~\cite{Shenvi03}. The hypercube is 
constructed by tensor (Cartesian) products of two-site linear graphs. The 
eigenvectors are therefore tensor products of the two solutions for the 
two-site graph. These solutions are represented by the Hadamard matrix 
$\overline{H}$. The Hamiltonian and its eigenvectors for the quantum wires 
(the collapsed hypercubes) are therefore contractions of the associated 
hypercube quantities.

It should be pointed out that the relationship between Krawtchouk functions and
spectral transforms has been discussed previously in the literature, primarily
in the context of optical and digital filtering~\cite{Atakishiyev97,Yap03}. 
These ideas were extended to quantum mechanical systems, where the Krawtchouk
transform can be considered as a fractional Fourier 
transform~\cite{Hakioglu00}. That said, the idea presented in this paper of 
performing the spectral tranform by time-evolution under the influence of a 
local operator is entirely new.

It is also important to mention that the discrete nature of the Hamiltonian is
a crucial ingredient that enables the spectral transform to be performed 
simply by turning on a linear potential. Suppose we attempted the same 
approach in the continuum limit $N\to\infty$ in which the Krawtchouk functions
reduce to Hermite polynomials. In this limit, the inhomogeneous tight-binding
Hamiltonian is replaced by a harmonic oscillator Hamiltonian with a potential
$V(z)={1\over 2}m\omega^2z^2$ with $\omega=2J/\hbar$ as discussed at 
the end of Sec.~\ref{sec:H0}. Adding a linear potential $V(z)=2J[z/(\lambda/2)]
=4Jz/\lambda$, the continuum limit of the discrete linear potential
$\tilde{V}_j=2j$ required to effect the spectral transform, the total potential
becomes 
\begin{equation}
V(z)={1\over 2}m\omega^2\left(z+{4J\over m\omega^2\lambda}\right)^2
-{8J^2\over m\omega^2\lambda^2}.
\end{equation}
The linear potential simply shifts the origin of the effective oscillator
potential to the left, so the eigenstates of the unperturbed Hamiltonian are 
instantaneously placed at the classical turning points of the shifted 
potential. The subsequent evolution is simple harmonic motion, with all
wavepackets initially propagating to the left. Under no circumstances will 
certain eigenstates propagate to the right!

One more note should be made before discussing the feasibility of performing 
this spectral transform in actual optical lattices in the next section. The
potential that must to be applied ($V(j)=2Jj$) in order to effect the 
spectral transform corresponds exactly to the eigenvalues of the original 
(unperturbed) Hamiltonian, listed along the diagonal in increasing order.
Curiously, this was exactly the same criterion found to satisfy the condition
(\ref{condition}) for the uniform lattice, discussed at the end of 
Sec.~\ref{sec:transform}, though in that case the full spectral transform 
could not be implemented because the resulting eigenvalue spectrum was not an
integer power-law. This observation could provide a tantalizing clue to 
generalizing the spectral transform by Hamiltonian evolution for other 
discrete systems.

\section{Approximations and Interactions}
\label{sec:approx}

The implementation of a spectral transform by Hamiltonian evolution described 
in the previous section assumed that there is only one particle in the entire 
lattice. This is of course not generally the case in real experiments, where 
there are many atoms and collisions between pairs of particles is not 
negligible. It is therefore important to investigate the efficacy of the 
approach including the effects of particle interactions. This is accomplished 
by including the interaction term proportional to $U$ in the 
Bose-Hubbard Hamiltonian~(\ref{Hubbard}), which has been so far neglected.

Likewise, it has been assumed that the site-dependent tunneling coefficients
$J_{i,i+1}$ in (\ref{wire}) could be implemented in actual experiments. In
practice, the tunneling amplitudes can be made to vary continuously over the
length of the lattice, as discussed in Sec.~\ref{sec:optlat}, but this 
variation is unlikely to yield exactly the correct values of the $J_{i,i+1}$.
This section considers separately the robustness of the spectral transform
when the exact conditions discussed in Sec.~\ref{sec:transform} are no longer
satisfied.

\subsection{Quadratic Optical Lattice}
\label{quadratic}

The exact tunneling coefficients (\ref{wire}) required for the spectral 
transform would not be straightforward to implement experimentally. As 
discussed in Sec.~\ref{sec:optlat}, a natural variation of the lattice
depth arises from the laser beam waists $z_0$, $V(z)=V_0\exp(-z^2/z_0^2)$. 
With the substitution $z=(\lambda/2)i$, the axial variation for the discrete
case is $V_i=V_0\exp[-(\lambda/2z_0)^2i^2]$. Inserting this into the 
tight-binding expression for the hopping coefficients~(\ref{tunneling}) and 
assuming that $z_0\gg\lambda$ one obtains
\begin{equation}
{J_{i,i+1}\over E_R}\approx
J_0\left[1+\left({\lambda\over 2z_0}\right)^2
\left(\sqrt{\overline{V}_0}-{3\over 4}\right)i^2\right],
\label{Jact}
\end{equation}
where $J_0=\sqrt{16\over\pi}\overline{V}_0^{3/4}e^{-2\sqrt{\overline{V}_0}}$ 
and terms of order $(\lambda/2z_0)^4$ have been neglected. Assuming 
$z_0=100~\mu$m, $\lambda=0.8~\mu$m, and $\overline{V}_0=30$, this quadratic 
approximation gives $J_{i,i+1}/E_R\approx J_0(1+7.5\times 10^{-5}i^2)$. Can 
such a quadratic variation of the tunneling coefficients reproduce the desired 
values~(\ref{wire})?

In the vicinity of the lattice center $i=(N-1)/2+j$ with $j$ a small integer 
$j\ll N$, the tunneling coefficients (\ref{wire}) are indeed approximately 
quadratic:
\begin{eqnarray}
J_{j,j+1}&=&J\sqrt{\left({N+1\over 2}\right)^2-j^2}\label{Jdes0} \\
&\approx&{N+1\over 2}J\left(1-{2j^2\over(N+1)^2}\right),
\end{eqnarray}
neglecting terms of order $\mathcal{O}(j^4)$. While this is an excellent
approximation near the lattice center, the tunneling coefficients decrease more
quickly near the lattice edges; clearly, if $j=(N-1)/2$ in the expression 
above one obtains $J_{(N-1)/2,(N+1)/2}\approx N/4$, which is much larger than 
the exact tunneling coefficient $J_{(N-1)/2,(N+1)/2}=\sqrt{N}$. An alternative 
approximation that reproduces the desired values at both $j=0$ and 
$j=\pm(N-1)/2$ would be 
\begin{equation}
J_{j,j+1}={N+1\over 2}J\left(1-{4j^2\over(N+1)(\sqrt{N}+1)^2}\right).
\label{Jdes}
\end{equation}

Comparison of Eqs.~(\ref{Jact}) and (\ref{Jdes}) reveals a problem, however:
the coefficients of the site-dependent term have opposite signs. This is 
because the actual tunneling coefficients for a focused laser beam (\ref{Jact})
are smallest when the lattice depth (laser intensity) is greatest. In contrast,
the desired coefficients (\ref{Jdes}) are supposed to be largest at the
center, so that the lattice depth must be shallowest at this point. 

\subsubsection{Axially displacing lattice beams}

In order to prepare an optical lattice with the desired properties, one could 
axially displace the lattice beams relative to the each other, so that the 
intensity maxima are located a distance $d\gg z_0$ apart. The lattice 
potential would then be $\overline{V}(z)={1\over 2}\overline{V}_0
(e^{-z^2/z_0^2}+e^{-(z-d)^2/z_0^2})$. Written in terms of the new variable 
$x=z-d/2$ centered
midway between the two maxima, the potential becomes
\begin{equation}
\overline{V}(x)=\overline{V}_{d/2}e^{-x^2/z_0^2}\cosh\left({xd\over z_0^2}
\right),
\label{newpot}
\end{equation}
where $\overline{V}_{d/2}=\overline{V}_0e^{-d^2/4z_0^2}$ is the zeroth-order 
potential depth at the lattice center equidistant from the centers of the two 
beam maxima. Expanding about $x=0$ gives an approximate quadratic envelope
\begin{equation}
\overline{V}(x)\approx\overline{V}_{d/2}+\overline{V}_{d/2}{x^2\over 2z_0^2}
\left({d^2\over z_0^2}-2\right)
\label{quadpot}
\end{equation}
corresponding to tunneling coefficients
\begin{equation}
J(x)\approx J_{d/2}\left[1-{x^2\over 2z_0^2}\left({d^2\over z_0^2}-{1\over 2}
\right)\left(\sqrt{\overline{V}_{d/2}}-{3\over 4}\right)\right],
\label{2beams}
\end{equation}
where
$J_{d/2}=\sqrt{16\over\pi}\overline{V}_{d/2}^{3/4}
e^{-2\sqrt{\overline{V}_{d/2}}}$.
Comparison of (\ref{Jdes}) and (\ref{2beams}) immediately gives
$J=2J_{d/2}/(N+1)$ and 
\begin{equation}
{\lambda^2\over z_0^2}\left({d^2\over z_0^2}-{1\over 2}\right)
\left(\sqrt{\overline{V}_{d/2}}-{3\over 4}\right)={32\over(N+1)(\sqrt{N}+1)^2},
\end{equation}
where the substitution $x=(\lambda/2)j$ has been made. Of course, to enforce 
the hard-wall boundary condition, end caps~\cite{Strecker02,Meyrath05} must 
also be applied transversally at the optical lattice amplitude maxima located 
at $z=0$ and $z=d$.

According to Eq.~(\ref{e0}), the spatial variation of the potential 
depth~(\ref{quadpot}) would yield a site-dependent energy shift
that is not included in the assumed hopping Hamiltonian~(\ref{H0}). Using the
harmonic oscillator approximation to the Wannier functions yields
$\varepsilon(x)/E_R\approx\sqrt{\overline{V}_{d/2}}x^2d^2/4z_0^4$ ignoring the 
constant offset energy $\sqrt{\overline{V}_{d/2}}$. It is therefore necessary 
to manually apply an inverted parabolic potential, centered at the depth 
minimum, that can serve to cancel this effective quadratic potential energy 
contribution.

Let's consider what these results imply for actual experiments, assuming 
$^{87}$Rb atoms confined in a red-detuned optical lattice with wavelength
$\lambda=0.8~\mu$m. With $V_0=50E_R$ and $z_0=100~\mu$m, one could obtain a
lattice with $N=201$ sites choosing $d\approx 255~\mu$m. The potential depth
at the midpoint between beam maxima is $V_{d/2}\approx 10E_R$, deeply within 
the tight-binding limit. The coupling constant is then calculated to be 
$J\approx 2.3\times 10^{-4}E_R$, though the actual coupling constants vary
from $0.01E_R$ at the edge to $0.02E_R$ at the midpoint between laser maxima.
An external potential must be applied at all times, 
$\overline{V}_{\rm e}^0(x)\approx -5.1(x/z_0)^2$ so that the magnitude of the 
external potential reaches almost one recoil energy at the lattice edges.
Applying the additional linear potential ${V}_{\rm e}^1(x)=4J(x/\lambda)$
then allows for the implementation of the spectral transform in a time
$t_s=\hbar\pi/2\sqrt{2}J\approx 210$~ms.

Likewise, for a lattice with $N=301$ sites and $V_0=20E_R$, one requires 
$d\approx 223~\mu$m which yields $V_{d/2}\approx 5.8E_R$. The external 
quadratic potential that needs to be applied is 
$\overline{V}_{\rm e}^0(x)\approx -3(x/z_0)^2$. With $J=4.6\times 10^{-4}E_R$ 
(the actual coupling constants vary between $0.036E_R$ and $0.069E_R$), the
time required is faster than for the example above, $t_s=107$~ms. It might at
first glance appear paradoxical that the $N=301$ case should be faster than 
the $N=201$ case, but recall that the total time $t_s$ is set by the parameter 
$J$, which for the latter case is twice as large as the value for the former 
case.

In order to verify that the quadratic optical lattice is indeed able to 
perform the spectral transform, an explicit numerical simulation of the time
evolution was performed. The governing tight-binding Hamiltonian is the 
unperturbed $H_0$ defined in Eq.~(\ref{H0}), but with the approximate 
tunneling coefficients $J_{i,j}=J(j)=J(x)$ given in Eq.~(\ref{2beams}); its 
eigenvalues and eigenvalues and eigenvectors are $\epsilon_k$ and 
$|v_k\rangle$, respectively. For simplicity, the initial state was chosen to 
be a Boltzmann distribution 
\begin{equation}
\rho=\sum_k\exp(-\epsilon_k/k_BT)|v_k\rangle\langle v_k|,
\end{equation}
where $\rho$ is the thermal density matrix, $T$ is the temperature in Kelvin, 
and $k_B\approx 1.38\times 10^{-23}$~J/K is Boltzmann's constant. The 
time-evolution of the density matrix is obtained in principle by numerically 
integrating the equation of motion 
$i\hbar\partial\rho/\partial t=[H_0+4Jx/\lambda,\rho]$ where $[,]$ 
denotes the commutator and the correct linear potential has been explicitly 
added. Equivalently, because the particles are assumed to have no interactions,
one may simply allow each eigenvector of the unperturbed Hamiltonian to evolve 
separately under the influence of the Schr\" odinger equation, and then add 
their contributions to obtain the time-dependent density matrix
\begin{equation}
\rho(t)=\sum_k\exp(-\epsilon_k/k_BT)|v_k(t)\rangle\langle v_k(t)|.
\end{equation}
The spatially dependent particle density is then readily obtained using the 
relation $n(j,t)=\langle j|\rho(t)|j\rangle$ or 
$n(x,t)=\langle x|\rho(t)|x\rangle$.

\begin{figure}[t]
\begin{center}
\epsfig{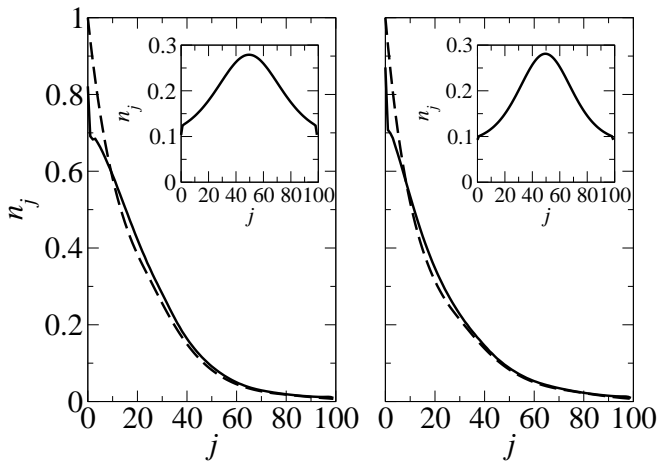}
\end{center}
\caption{Numerical results for the spectral transform. The parameters for the
simulation are $k_BT=40~E_R$, $z_0=100~\mu$m, $V_0=50~E_R$, and $N=100$. The 
left and right graphs correspond to $d=200~\mu$m and $300~\mu$m, respectively.
The results for the perfect tunneling coefficients (\ref{wire}) are given by 
dashed lines, while those based on tunneling coefficients (\ref{tunneling}) 
derived from the actual lattice potential (\ref{newpot}) are given by solid 
lines. The insets depict the density profile of the initial thermal 
distribution for the actual lattice.}
\label{sim1}
\end{figure}

The numerical results for the representative set of parameters $k_BT=40~E_R$, 
$z_0=100~\mu$m, $V_0=50~E_R$, and $N=100$ are shown in Fig.~\ref{sim1}, for
the choices $d=200~\mu$m and $300~\mu$m. Two cases were considered. In the 
first, the tunneling coefficients were chosen to be the exact ideal tunneling 
coefficients~(\ref{wire}). As expected, the
initial thermal distribution over energies is exactly reproduced after a time
$t=\pi/2\sqrt{2}J$ as a spatial distribution over lattice sites $j$, shown as
the dashed lines in the figure. In the second case, the tunneling coefficients 
were obtained for the actual lattice by inserting the spatially dependent 
optical lattice potential~(\ref{newpot}) into the general 
expression~(\ref{tunneling}) 
appropriate in the tight-binding limit. The results are shown as solid lines
in the figure. It is clear that for both values of $d$, the actual optical
lattice does an excellent job of reproducing the spectral transform. This
indicates that a careful adjustment of the separation of the intensity
maxima for the two optical lattice lasers (subject to $d\gg z_0$) is not 
necessary to ensure a successful implementation of the spectral transform. 
These general conclusions are also found to be independent of temperature.

It is also worthwhile to point out that the ability to perform the spectral
transform directly in the optical lattice greatly simplifies both the 
measurement of temperature and the ability to cool the atoms. The initial
particle density (shown as insets in Fig.~\ref{sim1}) is very broad, with
tails that are strongly truncated by the lattice edges. This would make a
direct measurement of temperature, obtained by fitting the tails to an
exponential function, quite unreliable. Furthermore, while one could attempt 
to cool the initial cloud directly by ejecting particles far from the lattice 
center, one would have little control over the actual energies that were 
removed from the system.
In contrast, the particle density after the time propagation is almost an
exact spatial map of the initial energy distribution. The exponential tail
(at least for $k_BT<V_0$) is well-resolved, allowing an excellent determination
of temperature. Furthermore, evaporative cooling on this state simply requires
the removal of all atoms occupying sites with index $j\gtrsim 1$, followed by
another spectral transform back to the energy basis.

\subsubsection{Spatial variation by interference}

While the approach described above of offsetting the focal distances of the 
two laser beams is able to reproduce the desired spectral transform, many 
experiments are performed under the assumption that the hopping amplitudes are
approximately constant over the region of interest. It would be preferable to 
employ a homogeneous optical lattice (with hard-wall boundary conditions) for
experiments, and then to only introduce the spatial inhomogeneity when 
temperature measurements or cooling is desired. Unfortunately, changing the 
focal length in the middle of an experiment is not feasible.

An alternative approach to modulate the optical lattice is through 
interference. Suppose that the homogeneous lattice is formed from two
counter-propagating laser beams each with amplitude $A$ and frequency $\Omega$
in the standard lin-$\perp$-lin configuration, i.e.\ where the two 
polarizations $\vec{\epsilon}$ perpendicular to one another. The electric 
field is then
\begin{eqnarray}
\vec{E}(z)&=&A\epsilon_xe^{i2\pi z/\lambda}e^{-i\Omega t}
+A\epsilon_ye^{-i2\pi z/\lambda}e^{-i\Omega t}\nonumber \\
&=&Ae^{i2\pi z/\lambda}e^{-i\Omega t}\left(\epsilon_x
+\epsilon_ye^{-i4\pi z/\lambda}\right).
\end{eqnarray}
The spatial variation of the polarization vector gives rise to a periodic 
atomic polarizability, which in turn yields the lattice potential 
$V(z)\propto\cos^2(2\pi z/\lambda)$. If one now introduces a second beam with
electric field amplitude $B\ll A$ oriented in the $yz$-plane at a small angle 
$\theta$ with respect to the $z$-axis,
\begin{eqnarray}
\vec{E}_2(z)&=&Be^{-i\Omega t}\Big\{\epsilon_x
e^{i2\pi[z\cos(\theta)+y\sin(\theta)]/\lambda}\nonumber \\
&+&\epsilon_ye^{-i2\pi[z\cos(\theta)+y\sin(\theta)]/\lambda}\Big\}
\end{eqnarray}
then the total lattice potential becomes
\begin{equation}
V(z)=V_0\left\{\cos\left({2\pi z\over\lambda}\right)
+\alpha\cos\left[{2\pi z\over\lambda}\cos(\theta)\right]\right\}^2,
\label{interpot}
\end{equation}
where $\alpha\propto|B|^2/|A|^2\ll 1$ is the relative depth of the second 
lattice beam.

For small angle $\theta\ll 2\pi$ so that $\cos(\theta)\approx 1-\theta^2/2$, 
Eq.~(\ref{interpot}) can be rewritten as
\begin{equation}
V(z)\approx V_0\left\{\cos\left({2\pi z\over\lambda}\right)
+\alpha\cos\left({2\pi z\over\lambda}-{2\pi z\over\lambda_b}\right)\right\}^2,
\end{equation}
where the `beat wavelength' has been defined as 
$\lambda_b\equiv 2\lambda/\theta^2\gg\lambda$, for reasons that will be clear
shortly. The potential then becomes
\begin{eqnarray}
V(z)&\approx&V_0\left[1+\alpha\cos\left({2\pi z\over\lambda_b}\right)\right]^2
\cos^2\left({2\pi z\over\lambda}\right)\nonumber \\
&\quad&+\alpha\sin\left({4\pi z\over\lambda}\right)\sin\left(
{2\pi z\over\lambda_b}\right).
\end{eqnarray}
If $\alpha$ is small then the last term in the above expression is small and
can be neglected. The potential is therefore described by a spatially varying
depth $V_0\left[1+\alpha\cos\left(2\pi z/\lambda_b\right)\right]^2$ which is
periodic at the beat wavelength. Unlike the case $\alpha=1$ often encountered
in physics, however, the potential depth never vanishes: the maximum depth 
$V_{\rm max}\approx V_0(1+2\alpha)$ corresponds to $z\approx 0$, while the 
minimum depth for $z\approx\lambda_b/2=\lambda/\theta^2$ is 
$V_{\rm min}\approx V_0(1-2\alpha)$. 

Expanding about the minimum 
$z=x+\lambda_b/2$ in the regime $x\ll\lambda_b$ one obtains
\begin{equation}
\overline{V}(x)\approx\overline{V}_0\left[1-2\alpha
+{4\pi^2\alpha\over\lambda_b^2}x^2\right]\cos^2\left({2\pi x\over\lambda}
\right),
\end{equation}
where as usual $\overline{V}\equiv V/E_R$. Thus, again the lattice depth is 
approximately a quadratic function of the lattice position. This leads to a 
quadratic site-dependent energy contribution~(\ref{e0}):
\begin{equation}
\overline{\varepsilon}(x)\approx\sqrt{\overline{V}_0\over 1-2\alpha}
{2\pi^2\alpha x^2\over\lambda_b^2}
\label{e0inter}
\end{equation}
neglecting site-independent energies.
Substituting this expression into Eq.~(\ref{tunneling}) yields the
site-dependent tunneling coefficients in vicinity of the depth minimum
\begin{equation}
J(x)\approx J_{\lambda_b/2}\left[1-4\alpha\pi^2{x^2\over\lambda_b^2}
\left(\sqrt{V_0}-{3\over 4}\right)\right],
\end{equation}
where 
\begin{equation}
J_{\lambda_b/2}=J_0\left[1+2\alpha\left(\sqrt{V_0}-{3\over 4}\right)\right]
\end{equation}
is the tunneling amplitude at the lattice minimum in terms of the bare
tunneling amplitude $J_0$.

Comparison with Eq.~(\ref{Jdes}) gives $J=2J_{\lambda_b/2}/(N+1)$ and
\begin{equation}
{4\over(N+1)(\sqrt{N}+1)^2}=\alpha\left({\pi\lambda\over\lambda_b}\right)^2
\left(\sqrt{V_0}-{3\over 4}\right).
\label{intercond}
\end{equation}
It is again useful to insert numbers that would be relevant to experiments with
ultracold atoms. With $\theta=\pi/25$, the beat wavelength is 
$\lambda_b=126.6\lambda$ or approximately 253 lattice spacings. One should
therefore be able to effect a spectral transform with $N\lesssim 201$. Choosing
$V_0=10~E_R$ and $N=101$, condition~(\ref{intercond}) yields 
$\alpha\approx 0.216$, giving $J\approx 9\times 10^{-4}~E_R$ and therefore
$t_s\approx 54$~ms. Note that an inverted parabolic potential 
$\overline{V}_{\rm e}^0(x)\approx -18(x/\lambda_b)^2$ must be applied at all 
times to cancel the site-dependent energy contribution~(\ref{e0inter}).
Likewise, choosing $N=201$ requires $\alpha\approx 0.058$ 
and $t_s\approx 170$~ms. For larger values of $N$, one would require smaller
values of $\theta$ to ensure that $\lambda_b\gg N(\lambda/2)$.

The main advantage of this approach is that temperature measurements and/or
cooling can be effected at any stage of an experiment with ultracold atoms in
optical lattices: one simply adds the second lattice beam and the appropriate
quadratic and linear potentials. Of course, these must be added adiabatically
in order to avoid heating the system in the process. A conservative estimate
of the adiabaticity timescale would correspond to the time taken by a
wavepacket to propagate fully across the lattice a few times, 
$t_{\rm adiab}\approx 2\hbar/J$ from Eq.~(\ref{sitespan}). For the $N=101$ 
example above, one obtains $t_{\rm adiab}\approx 100$~ms so that a ramp-up 
of the additional beam and the external quadratic potential over $200$ to 
$300$~ms would ensure that the system would be minimally perturbed.

\subsection{Effect of Interactions}

Until now the interaction term in the Bose-Hubbard Hamiltonian~(\ref{Hubbard})
has been completely neglected. In principle, the strength of particle 
interactions can be adjusted close to zero through the use of Feshbach
resonances~\cite{Fattori08}, but in general the gas will not be perfectly 
ideal. Unfortunately, a full numerical simulation of the spectral transform,
as was discussed in the subsection above, is not straightforward in the 
presence of interactions. The interactions lead to mixing among the energy
levels during the time propagation that are impossible to model in an average
way without introducing additional assumptions.

Instead, the spectral transform was numerically simulated by exact propagation
of the many particle Hamiltonian, but for a smaller number of lattice sites
$N=10$. The tunneling coefficients corresponded to the ideal 
expression~(\ref{wire}), and the values of $U$ in Eq.~(\ref{Hubbard}) and the
number of particles $1\leq N_p\leq 10$ were varied independently. The largest
number $N_p=10$ for $N=10$ corresponds to a Hilbert space dimension of 
$(N+N_p-1)!/(N_p-1)!N!=92\,378$; the computational time for the simulation of 
the spectral transform became unpleasantly long for larger $N_p$. The goal was 
to determine the fidelity $F$ of the spectral transform; i.e.\ if the initial
state was the $k$th eigenvector of the unperturbed (but interacting) 
Hamiltonian, then $F$ is the particle density in the $k$th site of the lattice
at the time $t_s$.

\begin{figure}[t]
\begin{center}
\epsfig{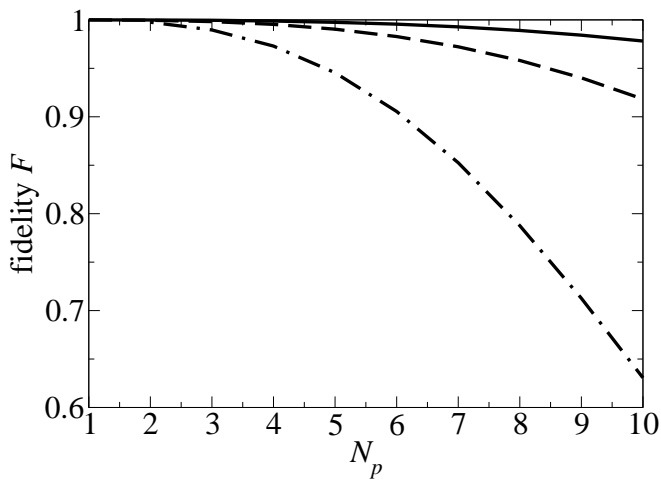}
\end{center}
\caption{Fidelity $F$ as a function of the number of particles $N_p$ obtained 
by exact numerical simulation of the spectral transform. With the initial state
the ground state of the unperturbed interacting Hamiltonian, the fidelity is 
defined as the density in the first site at time $t_s=\pi/2\sqrt{2}J$. The
solid, dashed, and dot-dashed lines correspond to $U=0.1J$, $0.2J$, and 
$0.5J$, respectively.}
\label{sim2}
\end{figure}

In practice, one need only consider the fidelity of the 
$|v_0\rangle\leftrightarrow|0\rangle$ spectral transformation, because this is
representative of all the eigenstates. The results of the numerical simulation
are shown in Fig.~\ref{sim2}. When $U=0.1J$, the fidelity has dropped to only
$0.98$ when $N_p=N=10$, i.e.\ at unit lattice filling $\nu\equiv N_p/N=1$. The 
best fit to the data indicates that the fidelity drops as a power law, 
$F\approx 1-0.023\nu^{3.3}$. As the on-site interaction strength is 
increased the fidelity drops; when $\nu=1$ one obtains $F\approx 0.92$ when
$U=0.2J$ and only $F=0.63$ when $U=0.5J$. The best fits in the $U=0.2J$ and
$U=0.5J$ cases are $F\approx 1-0.088\nu^{3.3}$ and $F\approx 1-0.43\nu^{3.1}$,
respectively, both power laws with a similar exponent. Though it is impossible 
to confidently extrapolate to large system sizes with $N$ and $N_p$ comparable 
to those in a future experiment, the numerical results suggest that the ability
to perform a spectral transform to 90\% accuracy or better would require 
$\nu\lesssim 0.6$ and/or $U\lesssim 0.5J$. While this value of $U/J$ ensures 
that the system is in the superfluid regime~\cite{Stoferle04}, it is smaller 
than the value one would generally obtain in the 1D lattice case (cf.\ 
Sec.~\ref{sec:optlat}).

\section{conclusions}
\label{sec:conclusions}

The scheme described above to perform the spectral transform with ultracold
atoms has assumed that the optical lattice is strictly one-dimensional. While 
this assumption has simplified the mathematical analysis and the numerical 
simulations, it is not necessary. In practice, many Bravais lattices are 
generated by two counter-propagating laser beams in different (not necessarily
perpendicular) spatial directions. One could envisage that the lattice beams
along each different axis would be suitably prepared so that their intensities
were varying quadratically, using either approach discussed in 
Sec.~\ref{sec:approx}. The tunneling amplitudes would then vary along each 
axis, so that the eigenstates of the lattice (in the non-interacting limit) 
would correspond approximately to the product of Krawtchouk functions for each 
direction. 

Performing a full three-dimensional
spectral transform would not necessarily require a linear potential applied in 
each direction simultaneously, a task which might be difficult in practice. 
Instead, one could simply apply the linear ramps for each dimension in 
sequence, separately cooling the atoms' motions in each spatial degree of 
freedom. For the cubic lattice for example, eliminating atoms from large site 
values in one direction would not perturb the atoms in the perpendicular 
directions, and cooling can be efficiently performed. The presence of 
interactions will modify this picture somewhat, but if the interaction 
strength is kept at sufficiently small ($U\lesssim J$ based on the 
one-dimensional simulations) then this should not present much of a problem.

Besides the obvious benefits of temperature measurement and cooling, the 
ability to perform a spectral transformation {\it in situ} has other possible 
future applications. For example, the technique could be used to probe the
excitation spectrum and heating mechanisms of strongly interacting systems. 
At any given moment, the interactions could be quickly tuned to approximately 
zero through the application of Feshbach resonances, and the system's energy
spectrum monitored by spectral transform followed by non-destructive imaging;
subsequently the system would be returned to its original state by inverting 
the process. Of course, care would need to be taken to ensure that the ramp on
and off of the interactions were slow enough that this process itself would not
greatly perturb the system. 

As a second example, it is conceivable that the technique could 
be extended to produce a greater variety of overall phases than the $0$ and 
$\pi$ in the present case~(\ref{condition2}). In particular, if a suitable 
choice of tunneling parameters and external potential could be found such that 
the resulting phases would be of the form $\pi km/N$, then the spectral 
transform would be equivalent to a quantum Fourier transform, a central 
operation in several efficient quantum algorithms including 
factoring~\cite{Browne07,Yoran07}. In this case, each lattice site (out of a 
total $N=2^n$) would label a possible state in the $n$-qubit Hilbert space. 
Though inefficient in terms of resources, this notion could pave the way 
toward implementation of interesting quantum algorithms using ultracold atoms 
in optical lattices with minimal external controls.

\begin{acknowledgments}
It is a pleasure to thank M.~Oberthaler for stimulating conversations. This 
work was supported by the Natural Sciences and Engineering Research Council of
Canada and the Canada Foundation for Innovation.
\end{acknowledgments}

\end{document}